# The beginnings of celestial navigation: early techniques and instruments

**Gabriele Vanin**
**Independent scholar (rheticus@tiscali,it)**

**Abstract.** The use of the observed positions of celestial bodies to determine a navigator's location and to direct vessels, was an aspiration of ancient seafarers. Various peoples, in the Mediterranean as much as in the Indian Ocean, in China as much as in the Pacific, looked up to the sky as a guide for ships, but there is no convincing evidence this was achieved until the Modern Age. It was not before the Modern Age that two techniques were developed by the Portuguese, the measurement of the North Star's altitude and that of the Sun's meridian altitude , which, adapted and simplified for the use of sailors, enabled them to achieve good accuracy at least as far as the measurement of latitude was concerned. To this end, instruments that had already been used by astronomers for several centuries were also adapted and simplified: the quadrant, the astrolabe and, later, the cross-staff.

## Introduction

Exactly 500 years ago, on 6th September 1522, the most extraordinary adventure of exploration on planet Earth was accomplished: the *Victoria*, commanded by Juan Sebastian Elcano, the only ship left out of the five (with 237 men on board) of Magellan's fleet that had left Seville three years earlier, entered the bay of San

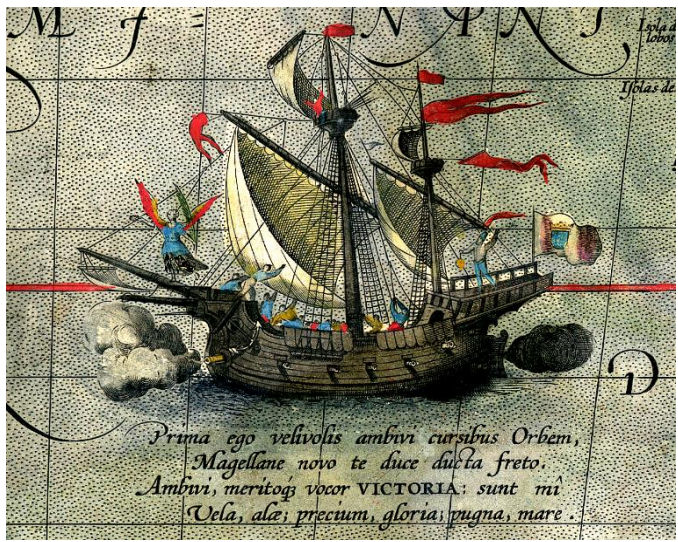

Lucar de Barrameda. The 18 men on board, the only survivors, were the first to circumnavigate the Earth, covering 14 460 leagues,[1] or 80 600 km. The voyage was of paramount importance from a geographical point of view, chiefly because it was the most striking experimental verification of the Earth's roundness. It also proved that our planet was actually larger than Ptolemy had believed, a fact giving credence to earlier measurements made by Eratosthenes, that the Pacific was much wider than it was thought, that America was as circumnavigable as Africa, and that if you go around the globe you lose a day's time when going west, whereas when going east (like Jules Verne's Phileas Fogg),[2] you gain a day's time.

It is generally believed that the role played by celestial navigation methods in promoting this and the other great geographical explorations carried out between the 15th and 16th centuries was prominent, but reality, as we will try to prove, is slightly more complex.

**Fig. 1.** *Detail of a 1590 map by the Flemish cartographer Abraham Ortelius showing the ship* Victoria *which, under the command of Captain Elcano, was the first to sail around the world.*

---

**Antiquity**

If by celestial navigation we mean the observation of celestial objects as sights to know one's position on Earth and to steer one's vessel correctly towards the chosen destination, then it can perhaps be assumed that it was also somehow practised in the earliest times.

**The Greeks and the Phoenicians**

This is how Homer's reference can be interpreted, when Ulysses is instructed on the direction to take in order to cross from the island of Ogygia (whose location is uncertain, but by all means west of the Greek hero's homeland) to Ithaca:

> *…nor e'er did slumber fall upon his eyelids,*
> *still on the Pleiads gazing, and Boötes,*
> *late-setting, — and the Bear, the Wain by others*
> *surnamed, which makes its round, Orion watching,*
> *and, sole of stars, ne'er dips in Ocean's waters:*
> *for it Calypso, nymph divine, enjoined him*
> *to keep in sailing to the left hand ever.*[3]

A further source is Aratus, who provides many references to maritime practice and where the River (Eridanus) and Orion are specifically mentioned as night markers:

> *On a clear main the sailor now may note*
> *the Stream's first bend arising from the deep,*
> *whilst waiting for Orion to supply*
> *some marks of darkness' length or of his voyage.*[4]

and the two Bears for orientation purposes:

> *                …by it on the deep*
> *Achaians gather where to sail their ships;*
> *Phoinikians to her fellow trust at sea.*
> *Twister is clear and easy to perceive,*
> *shining with ample light when night begins;*
> *though small the other, 'tis for sailors better,*
> *for in a smaller orbit all revolves:*
> *by it Sidonians make the straightest course.*[5]

Similar notions about the stellar preferences of the Achaeans and the Phoenicians, evidently borrowed from Aratus (which confirms the great influence of the distinguished poet on his later successors) are found in other ancient authors, such as Ovid[6] and Valerius Flaccus.[7]

As a matter of fact the Phoenicians, unlike the other Mediterranean peoples who mainly practised cabotage, had direct shipping routes across the Mediterranean, on which they reached Cadiz as early as 1100 BC[8]. Under the orders of the Egyptian pharaoh Necho II they circumnavigated Africa[9] around 600 BC and were likely to have reached the Canary Islands, Madeira and the Azores. Remarkably Thales, a Phoenician by origin,[10] is credited with the first known astronomical treatise, *Nautical Star-guide*.[11]

---

[3] *Odyssey*, 5, 271-277; translation of Henry Alford, *The Odyssey of Homer* (London: Longman, Green, Longman and Roberts, 1861), pp. 84-85.

[4] *Phenomena*, 728-731, translation of Robert Brown, *The Phainomena or 'Heavenly displays' of Aratos* (London: Longmans, Green, and co., 1885), p. 67.

[5] Ibid., 37-44, pp. 15-16.

[6] *Fasti*, III, 107-108.

[7] *Argonautica*, 1, 16-18.

[8] Piero Bartoloni, "Fenici e Cartaginesi sul mare", *Le Scienze*, 130 (1979), 30-36, p. 30.

[9] Herodotus, *Histories*, 4, 42.

**The Romans**

It seems that the ancients were well aware that the most obvious method to find one's way at sea was to look at the sky. Virgil even went as far as entrusting sailors with the art of identifying and naming constellations:[12]

> *...then gave the sailor names and numbers to the stars:*
> *Pleiades, Hyades and the shining Lycaon's Bear;...*

And the most relevant reference to the practice of astronomical navigation comes again from the Roman world and is found in Lucan's *Pharsalia*, when Pompey enquires with a ship's captain which stars he uses to navigate, and the latter replies as follows:[13]

> *To all those who flow through the starry sky*
> *on the vault that never stands still and deceives the poor sailors,*
> *to the stars we pay no heed; but what does not plunge in the waves,*
> *the axis that does not set, conspicuous by the two Bears,*
> *governs the ships. When this on high for me always*
> *will rise, and the Ursa Minor will remain on the top of the antennas,*
> *the Bosporus and Pontus curving the coast of Scythia*
> *we shall see. When descends from the summit of the must*
> *Arctophylax and Cynosura is brought nearer to the sea,*
> *the ship will turn to the ports of Syria. Then Canopus*
> *intent on wandering in the southern sky, welcomes us,*
> *the star that fears Borea; keep it to the left*
> *beyond the Lighthouse, the ship in the middle of the sea will go to touch the Sirte.*

However fanciful and imaginative these indications may be, as they even contradict what is stated in the first lines, the reference to the position of the celestial pole ("the axis that does not set") as an indication to find the North is still correct. This confirms what Aratus reported, namely that the narrower path of Ursa Minor around the celestial pole enabled ancient seafarers to find the north more easily, when Polaris was far away from its present position. To what extent this operation was feasible is in fact difficult to imagine. However, all these instances do not provide any indication of the possible use of particular techniques or instruments for quantitative measurements. In the end, it does not appear that any such techniques existed. At least, nothing has come down to us.

**In the Pacific**

From 2000 BC, the Polynesians expanded and populated the whole of the Pacific, island by island starting from the chain of coastal islands in South-East Asia. They occupied in a row the Solomons (1600 BC), the New Hebrides, the Fiji, the Samoa (1200 BC), reaching the Marquesas around the first century, the Hawaii and Easter Island around the fifth century and New Zealand around 1000.[14]

These peoples' deep knowledge of the starry sky was already mentioned in 1769 by the Europeans following Captain Cook, who wrote how they used it for navigation[15] and, in more detail, by the Spanish Captain José de Andía y Varela, who was in Tahiti in 1774-75 and wrote:

---

When the night is a clear one they steer by the stars; and this is the easiest navigation for them because, these being many [in number], not only do they note by them the bearings on which the several islands with which they are in touch lie, but also the harbours in them, so that they make straight for the entrance by following the rhumb of the particular star that rises or sets over it; and they hit it off with as much precision as the most expert navigator of civilised nations could achieve.[16]

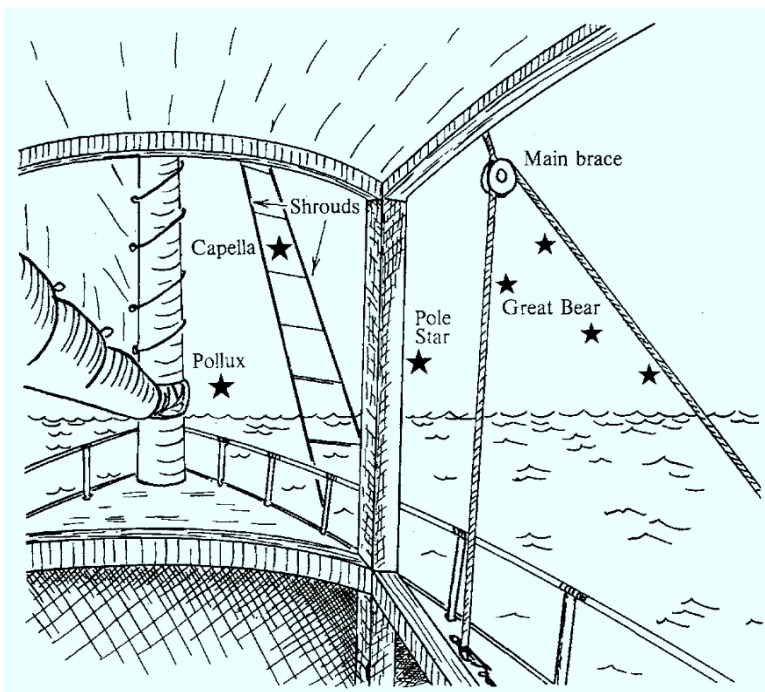

**Fig. 2.** *Some stars and asterisms used as "star paths" by Polynesian navigators (from Lewis,* We, the navigators*).*

Over time, anthropologists gradually realised that all the astronomical knowledge of the peoples of the Pacific, in Micronesia[17] as well as in the Gilberts,[18] in Tahiti[19] as well as in Tonga,[20] had been developed essentially for navigation, although of course it is not known how far back their knowledge dated.

Stars were used in a variety of ways. One way,[21] the "star at zenith" (*fanakenga*) was used to know one's latitude, based on the fact that at the moment of its culmination a star is directly at the zenith of that location having a latitude equal to its declination. It follows that an observer who notices a particular star directly above his head will know that his latitude will be equal to the star's declination. Thus, for example, when Sirius passed overhead, one was at the latitude of Tahiti and Fiji, when Altair passed one was at the latitude of South Marshall, when Arcturus passed one was at the latitude of Hawaii, and so on.[22] However, this practice is no longer in use anywhere in the Pacific.[23]

The method of the "star path" was used (*kaveinga* in Tonga,[24] *kavenga* in Tikopia in the Solomon Islands,[25] *aveia* in Tahiti[26]) to set the course towards a certain island i.e. a path of stars pointing towards that island, rising stars if the route was eastern, setting stars if it was western. Up to 10 stars were used in one night, rarely more. When the star had risen too high or dipped too low, the next star was followed at sunrise or sunset.[27] The planets[28] and the rising and setting points of the Sun[29] were also used to this purpose. Stars on the bow, but also on the stern and on the side, at any angle and in various positions would  be used  in relation to the equipment,[30] and the indications given by the stars could be corrected depending on  winds and currents.[31]

In the Caroline Islands a stellar compass was available, where the directions, 32 as in magnetic compasses, are indicated by the rising and setting points of 17 stars or star groups rather than by cardinal points (the Polaris and the Southern Cross placed vertically indicating north and south respectively). This stellar compass has no practical orientation purpose, but is rather an abstract representation providing spatial coordinates, in the sense that all directions in which one navigates are expressed not with respect to the cardinal points, but with respect to the rising or setting of these stars, and the directions from which winds or currents blow are also expressed in this way.[32]

No instruments of any kind seem to have ever existed in Oceania for this type of observation,[33] and the only documented traces of measurements of stellar altitudes are made by estimation. For example, the altitude of the Polaris is estimated by eye or by placing a span at arm's length: this measurement is an *ey-ass* (which identifies the hooked pole for harvesting breadfruit) or 15°. Lewis was told that Polaris was *half ey-ass* high at Satawal in the Carolinas, *one ey-ass* high at Saipan in the Marianas, which was substantially correct, and *one ey-ass and a half* high at Honolulu, about one degree too high.[34]

**The Middle Ages**

**The Vikings**
The Vikings, coming from Sweden, Denmark and Norway, were probably the greatest navigators in the Middle Ages . They carried out considerable exploration and colonisation in the North Atlantic between the ninth and tenth centuries.[35] Around 860 they reached Iceland, which was colonised by 930. In 981 they landed in Greenland, which was colonised five years later. Finally, around the year 1000, Leif Eirikson landed in an American territory called Vinland, between the estuary of St. Lawrence and New England, probably in Passamaquoddy Bay, at 45° north latitude, on the Canada-US border.[36]

Whether the Vikings used any kind of celestial navigation is doubtful. They are believed to have been able to navigate along the same parallel, perhaps using the approximate altitude of the Sun or Polaris. The scant half of a round wooden disc, dating from about 1200, was found in 1951 in a Benedictine nunnery in Greenland, at Uunartoq, in the Eastern Settlement. As it had notches on it, it was interpreted as an instrument combining the characteristics of a sundial and a direction finder.[37] This interpretation, however, was immediately met with a lot of perplexity and scepticism partly because of its small size (only 7 cm in diameter), it could in fact have been a much more common item, basically anything else.[38]

---

## The Arabs

The only Arabic texts available on nautical science were written rather late, between the end of the 15th century and the beginning of the 16th, by Ahmad ibn Mājid and Sulaymān al-Mahrī. They supposedly stand at the end of a tradition, centred on the Indian Ocean, of which they are descendants but for which we have no older sources. All we have is a few suggestions: apparently an Indian master, Aryā Sūra, who lived in the early centuries of our era, in a treatise called *Jātakamālā* listed what would have been the qualities of a good pilot as sought after by Indian merchants and there he mentioned "knowing the course of the celestial luminaries".[39]

In 1447 the Venetian merchant Niccolò De Conti wrote:[40]

> The sailors of India govern themselves by the stars of the Antarctic pole,... because they rarely see our Tramontana, and... they govern themselves according to the fact that they find the said stars, either high or low, and this they do with certain measures which they use,....

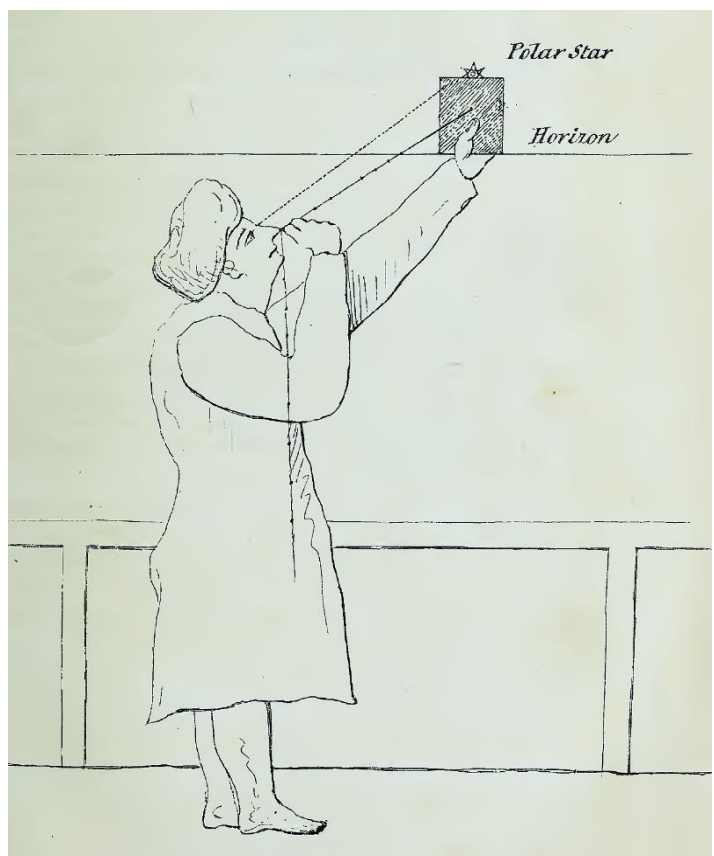

**Fig. 3.** *Method of use of kamal (from Congreve,* The Madras Journal of literature and science*).*

However, we have no evidence as to when this knowledge actually became scientific practice and if and when it eventually became the heritage of the Arab sailors who sailed across the Red Sea, the Persian Gulf and the Indian Ocean.

Ibn Mājid's main nautical text, the *Kitāb al Fawā'id*, was compiled in 1479-80 and it includes a series of treatises on various aspects of navigation, which, however, may sound rather obscure to us due to the scribes' numerous transcription errors. With regard to nautical astronomy, it illustrates, among other things, the use of the culmination of the 28 Islamic lunar houses for latitude measurements, alone or in combination with other stars and constellations. It also provides for the various lunar cases in culmination, the corresponding values of difference between the altitude of the Polaris and the celestial pole (*bāshī*); the text mentions stars and asterisms used for latitude measurements, in particular the Plough, the Guards and the Polaris; it gives tables of latitude measurements for various ports of the Indian Ocean taken using the Guards and the Polaris.[41]

The most important of Sulaymān al-Mahrī's five nautical texts, *'Umdat al-Mahrīya*, dating from 1511, is much simpler and more concise and it includes a theoretical part and a practical application immediately following, and is often helpful to understand some obscure passages in ibn Mājid; from an astronomic point of view it provides the most important part of the theory on the determination of latitudes from star altitudes, and a set of latitudes based on the altitude of the Polaris for all ports in the Indian Ocean.[42]

The *Kitāb al Fawā'id* mentions almost 70 methods for measuring the altitude of stars, but al-Mahrī explicitly states that only six or seven are needed under normal conditions. To ibn Mājid the most accurate method is the measurement of the meridian altitude of a star, its declination known; next comes the *abdāl* measurement, i.e. when two stars of equal declination appear at the same altitude on the horizon and form an isosceles triangle with a horizontal base with the pole, with many less accurate variations; other types of position include the vertical arrangement of two stars, or the measurement of the altitude of one star while another is at a certain altitude.[43]

The altitude of a star was measured in fingers (*iṣbaʿ*), i.e. the width of a finger held at the distance of an outstretched arm, which was between 1°36' and 1°43', and in *dhubbān*, i.e. four fingers side by side. One *dhubbān* was equivalent to the distance between α and β Aurigae.[44]

The instruments for altitude measurements are not described by Ibn Mājid and al-Mahrī, who speak generically of *khashabāt* ("pieces of wood"), but the Ottoman admiral Sidi Ali Çelebi describes two different versions in his 1554 work *al-Muḥīṭ*, a translation and rearrangement of the works by the two previous authors. The first is the one used by the ancients, which he calls *loḥ* ("tablet") consisting of nine rectangular tablets of different sizes, corresponding to different altitudes on the horizon, from four to 12 *iṣbaʿ*, strung on a length of string through a hole in the middle. The tablet was held with one hand and pointed towards the star to observe, while the end of the string was held in the other hand and close to the eye, so that the horizon was just below the lower edge of the tablet and the star just above the upper one. The second instrument, used at his time and called *ālat-al muqās*, consisted of a single rectangular tablet, with one side twice as long as the other, with a string passing through the centre and graduated with six knots corresponding to the various altitudes. The observation was made by holding the string between the teeth and moving the tablet closer or further away from the eye until it occupied the angular space between the celestial body and the horizon. The tablet could be used for both the long and short sides, which doubled the possibilities of measurement.[45]  An instrument similar to the first one described by Sidi Ali Çelebi seems to have been introduced to European navigators by the Arab pilot that Vasco da Gama employed while crossing the Indian Ocean in 1498:

> And as Vasco da Gama showed him the great wooden astrolabe which he had, and others made of metal, with which they took the altitude of the sun, the Moor was not surprised, saying that some pilots of the Red Sea used brass instruments of triangular shape and quadrants, with which they took the altitude of the sun, and especially of the star, and these were more widely used in navigation. But as their navigation was based on certain stars, both of the north and south, and other notable ones, which ran through the centre of the sky from east to west, he and the sailors from Cambay and all India did not take the distance with instruments similar to those [da Gama's]. He used a different tool instead, which was made of three tables, whose instruction he soon went on to show.[46]

Denis Papin described one device certainly similar to the second in a 1709 letter to Charles le Gobien, in which he describes its use in the Bay of Bengal:

> Their pilots take the altitude (or latitude of places) with a cord that has several knots in it. They put one end of the cord between their teeth, and by means of a piece of wood fixed to it, that has a hole through it, they easily observe the tail of *Ursa Minor*, which is commonly called the Polar Star, or north pole.[47]

James Prinsep happened to meet a pilot from the Maldives in the port of Calcutta who was still using a version of the second type of instrument, which he called *kámal*, and thus he describes it:

...a small parallelogram of horn (about two inches by one) with a string (or a couple of strings, in some instances), inserted in the centre. On the string are nine knots.[48]

The use of a similar instrument, which had knots corresponding to the latitudes of the various ports along the coast, was contemporaneously reported by sailors on the Coromandel coast.[49]

Specimens of this type, which Pereira da Silva also calls *kamal*, are preserved in the Ethnographic Museum in Hamburg: they were donated in 1892 by a captain who sailed on the Hamburg-Calcutta line and had belonged to a Hindu pilot  sailing on the Coromandel coast.[50]  They were also used by the Portuguese: Mestre João, Cabral's pilot, who called the instrument *las tablas de la índia*, experimented with it with modest success in 1500.[51] A further mention of  the use that the Portuguese made of the so-called *tavoletas* is to be found  in the *Livro de Marinharia* published by Brito Rebêlo in 1903 in Lisbon, which is a reproduction of a codex from the mid-16th century.[52]

It does not seem that Arab navigators of the time used the quadrant or the astrolabe, whatever de Barros and De Couto may say in the above quotation: ibn Mājid briefly mentions the astrolabe, but one gets the impression that he only knew it by name as the instrument used by astronomers to measure altitudes.[53]

**The Chinese**

In China, as in the West, there are very ancient, albeit vague, references to "the sea people"'s habit of orientation with stars. In particular, in *Huainanzi*, an ancient work from the second century BC, we read: "Those at sea who become confused and cannot distinguish east from west, orient themselves as soon as they see the pole-star".[54] Zhang Heng, the royal astronomer of the Han dynasty, author of one of the first star maps, wrote in his treatise *Ling xiàn*, published in 118 AD, that "there are in all 2500 stars, not including those which the sea people observe".[55]

The Chinese were taking stellar altitudes as early as the beginning of the 12th century. The *Kao-Li Thu Ching* (the report of a mission to Korea in 1124) states that pilots navigated using the Great Bear and nearby stars.[56]  A Song dynasty document (*Song Huiyao Jigao*) mentions warships equipped with sighting tubes used to determine the positions of the stars of the Great Bear.[57] The Southern Cross, Cassiopeia, Vega and stars of present-day Carena were also used as reference points. Some manuals referred to the rising and setting of constellations (Cassiopeia, Cross, Carena, Great Bear) as being useful for navigators to determine directions long before the invention of the compass, while others featured star charts for use in other countries. Tables with azimuth predictions of the rising and setting of the Sun and Moon were also in use for the same purpose.[58]

The last chapter of *Wubei Zhi*, a treatise on military and naval technology, printed in the early 16th century, includes maps and star charts relating to the voyages of Admiral Zeng He between 1400 and 1433. These provide instructions for holding to the positions of certain stars during certain routes, with measurements given in fingers and fractions of fingers, not in degrees, equivalent to the Arabic ones (a likely sign of contamination between the two cultures).[59] Although we do not generally know if and which instruments were used for measurements, at least in the earliest times, a detailed description of an instrument quite similar to the *kamal* can be found in the treatise *Chieh An lao Fen Man Pi*, written in the 16th century by Li Hsü and printed in 1606.[60]

**The Modern Age**

King Alfonso X of Castile (1252-1284) had ordered that all Spanish ships carry an astrolabe or quadrant to measure latitude, a fact that could be considered quite exceptional at the time because those instruments had not been adapted yet for use at sea.[61] However Ramon LLull, in 1272, reported that the Balearic sailors navigated following the North Star, and in 1299 he mentioned the astrolabe, saying that it was a useful instrument for sailors.[62]

Around the 13th century, necessity caused the gates of the Atlantic to open to European fleets. That was the time when the Arabs reconquered all the Syrian cities, snatching them from the crusaders. They also came to control trade over a vast region stretching from Egypt to India, imposing onerous tolls and effectively blocking European trade towards the East. Later, following Ming rule, the promising trade routes made known by Marco Polo's extraordinary voyage were also closed, and Tamerlane's tribes began to plunder Persia and the countries bordering the Black Sea, plunging Asia into anarchy and causing the Italians to lose access to the wealth of China and Central Asia.[63]

Thus began the progression along the African coast. In the 13th century, the navigation of the Atlantic was started by Italian sailors, mainly from Genoa. Between the 14th and 15th centuries, the Atlantic enterprise passed to the Portuguese, although at first, lacking their own local navy, they resorted to foreign admirals, mostly from Genoa.

The first champion of Portuguese exploration was the Infante Henry, known as "The Navigator". The prince, present in Lisbon between 1415 and 1433, promoted numerous expeditions along the African coast and to the Atlantic archipelagos.[64] Much has been written about the school of navigation, cosmography and astronomy supposedly founded by Henry in Sagres, but this is an unconfirmed myth, if only because he only settled in Algarve in 1452 and by the time of his death in 1460, only the perimeter walls of his fortified residence, which was to house the famous school, had been built.[65] In particular, astronomical navigation methods, as we shall see, were adopted after the Infante's death. In his time, and practically until the end of the 15th century, the so-called navigation by estimation was practised, even in the Atlantic, and sailors only relied on the assessment of the ship's speed and its direction, obtained with the compass.

**The instruments**

Towards the middle of the 15th century, pilots began to measure the altitude of the stars in order to obtain the latitude of the various ports and their current position. The earliest instruments used were the nautical quadrant, the nautical astrolabe, and later the nautical cross-staff.

## The nautical quadrant

The nautical quadrant has its ancient forerunner in Ptolemy's fixed quadrant,[66] which consisted of a piece of graduated wall placed on the meridian and equipped with a gnomon, whose shadow allowed the measurement of the Sun's altitude at noon. Many such instruments were designed by the Arabs[67] and they are likely to have transformed the quadrant into a smaller and more portable device with many other functions. The oldest description of such an instrument is the one contained in al-Battānī's tables (8th-9th century),[68] while in the West it is only mentioned in the first half of the 11th century.[69] The oldest extant specimens, all dating back to around 1300, are the one preserved in the Museum of the History of Science in Oxford, the one described by Dekker in 1995, and the one found in 2013 in the city of Zutphen in the Netherlands.[70] It is not known who first used it at sea, but the first recorded observation was made presumably from New Guinea by the Portuguese Diogo Gomes between 1460 and 1462 .[71] Gerard L'E Turner calls it altitude plain quadrant and says it had about 25 cm radius.[72] It does not appear that any ancient nautical quadrants, being mainly made of wood with only a few of brass, have survived.[73] Apparently, the first representation of the nautical quadrant only dates from 1528.[74]

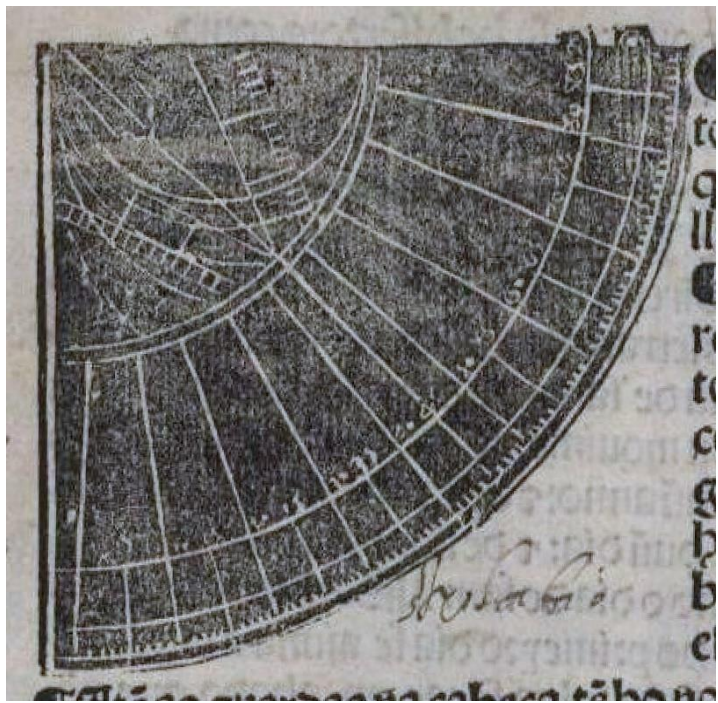

With the quadrant, the celestial body was viewed through two sights, with holes of about 3 mm in diameter, located on the upper segment of the instrument; the plumb line allowed the altitude of the celestial body to be read on the graduated scale. The instrument was kept suspended, if used on board, or fastened onto a support if used on land. Obviously, in less than calm seas it was difficult to use on board, but things improved if a second operator reading the altitude on the graduated scale was available. Like the astrolabe, and unlike the cross-staff or the modern sextant, which require the horizon to be visible at the same time, the quadrant could also be used in complete darkness, which was an advantage because it gave a better view of the stars. Even so, however, sighting a star like the Polaris, which is only of second magnitude, requires considerable skill, even on land, let alone at sea, since it is difficult to place the star at the centre of the farthest sight. In practice, you need a sky with a limiting magnitude no lower than fifth. Making the sights smaller does not serve the purpose, because holes of one to two mm in diameter literally make it impossible to see the stars. If built properly and used skilfully, this instrument is more accurate than one might imagine. Personally, with a self-made wooden quadrant of 26 cm radius and bright stars, I have obtained altitude measurements on the ground with an average error of a tenth of a degree, with the Polaris of a quarter of a degree.[75]

*Fig. 4. The first representation of the nautical quadrant (from* Repertorio dos tempos *by Valentim Fernandez).*

---

[66] *Almagest*, 1, 12.

[67] Jean-François Oudet, RégisMorelon, *Il ruolo e gli strumenti delle osservazioni astronomiche*, from *Storia della Scienza,* 3: *La civiltà islamica* (Rome: Istituto della Enciclopedia Italiana, 2001), pp. 172-181.

[68] Emanuel Poulle, *Gli strumenti astronomici e la misura del tempo*, in *Storia della Scienza*, 3: *Medioevo e Rinascimento* (Rome, Istituto della Enciclopedia Italiana), 2002, p. 523.

[69] Albert Anthiaume, Jules Sottas, *L'astrolabe-quadrant du Musée des Antiquités de Rouen* (Paris: Thomas, 1910), p. 73.

[70] Francis Maddison, **"**Medieval scientific instruments and the development of navigational instruments in the 15th and 16th centuries", *Revista da Universidade de Coimbra*, 24 (1968), pp. 13-14 and fig. 6; Elli Dekker, "An unrecorded medieval astrolabe quadrant from c. 1300", *Annals of science*, 52(1) 1995, 1–47; John Davis "The Zutphen quadrant: a very early equal-hour instrument excavated in the Netherlands", *The British Sundial Society Bulletin*, 26 (2014), 36-42.

[71] Diogo Gomes, "As relações do descobrimento da Guiné e das ilhas dos Açores, Madeira e Cabo Verde", *Boletim da Sociedade de Geografia de Lisboa*, 17(5), 1900.

[72] *Scientific Instrument 1500-1900: an introduction* (London: Wilson, 1998), p. 30.

[73] Alan Stimson, *The mariner's astrolabe* (Utrecht: Hes, 1988), p. 13.

[74] This is a drawing on f. g6v from a 1528 edition of Valentim Fernandez's *Repertorio dos tempos*.

## The nautical astrolabe

The nautical astrolabe would seem to derive from the planispheric astrolabe, whose verso was actually fully graduated on the edge and bore an alidade with sights, and could be used to measure the altitude of the stars. However, they are two conceptually different instruments, and the verso of the planispheric astrolabe actually derives directly from a different instrument described by Ptolemy and used to measure the obliquity of the ecliptic.[76]  The planispheric astrolabe was already known to Ptolemy, if not to Vitruvius and Hipparchus.[77]

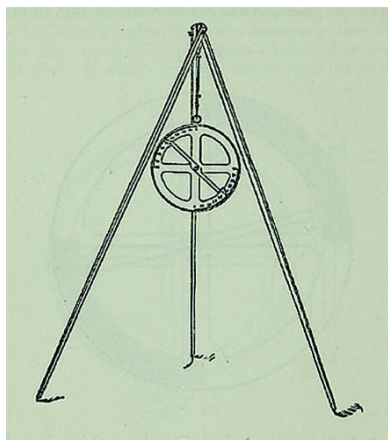

However, the first nautical astrolabes were actually similar to the back of the planispheric astrolabe, flat wooden or brass discs.[78]  It seems that the first sailor to use an astrolabe at sea,  during a 1481 voyage along the west coast of Africa, was the Portuguese Diogo de Azambuja, Captain Major of the army commissioned by King John II to build the fortress of São Jorge da Mina, in present-day Ghana.[79] It was also used by Bartolomeu Dias on his 1487-88 voyage,[80] by Vasco da Gama on his 1497 voyage,[81] and by one of Cabral's pilots, Mestre João, on his 1500 voyage.[82]  Apart from da Gama, it is likely that ordinary planispheric astrolabes were used for these observations,[83] partly because when Correa describes the astrolabe built by Zacuto, he does not seem to mean an instrument too different from these.[84]

When exactly the nautical astrolabe was introduced is not certain. The oldest description of one, which comes with a small sketch, is that given by the Venetian Alessandro Zorzi in a manuscript of 1517,[85] which mentions the bronze astrolabe already used by the Portuguese at that time in their navigations. The two main differences from the planispheric astrolabe are well described in it: the reduced distance between the pinnules was meant to make the movement of the alidade easier, reducing its momentum and easing its use on board ship. The second difference being the transformation from a complete disc into a perforated wheel, but of greater thickness and weight, which made the instrument more stable and less subject to the action of the wind. However, this evolution was probably rather gradual, or perhaps it slowly established in Spain. The earliest real illustration of a marine astrolabe is in the Castiglioni Planisphere, by Diogo Ribeiro (a Portuguese who entered the service of Spain in 1518). The picture bears the date 1525 right on the upper plate of the astrolabe,[86] and it still shows a complete disc, but with pinnules at a reduced distance. Even the first printed illustration and description of an astrolabe, and of its construction, graduation and use, by the Spaniard Martin Cortes,[87] published in 1551 but written in 1545,[88] refers to a disk astrolabe with pinnules close together. The first real picture showing the full transformation appears in Jean Rotz's *Booke of Hydrography*, a manuscript written between 1535 and 1542.[89]

*Fig. 5. Representation of how da Gama may have arranged his astrolabe to take the altitude of the Sun from the bay of St. Helena (from Pereira da Silva* O astrolábio náutico *in* Obras completas, 2*).*

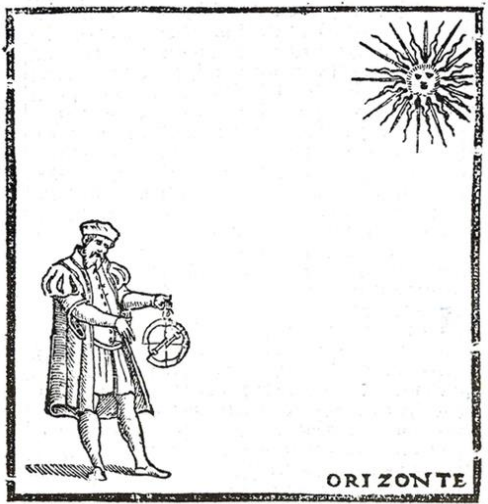

**Fig. 6.** *Use of the astrolabe to measure the altitude of the Sun (from de Medina*, Regimiento de navegación*).*

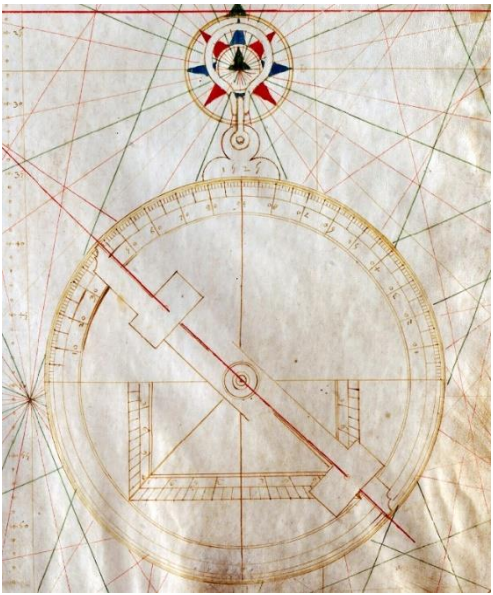

**Fig. 7.** *The astrolabe in the* Castiglioni Planisphere*.*

The three oldest extant nautical astrolabes were found in Namibia in 2008 in the wreck of the Bom Jesus, a Portuguese ship wrecked in 1533 but one is incomplete and the other two need restoration owing to heavy concretions.[90] The oldest complete and well-preserved nautical astrolabe, dated 1540, 200 mm in diameter and 15 mm thick, was in the National Museum in Palermo[91] but disappeared at the end of the Second World War.[92] The oldest complete and well-preserved still in existence is the Father Island III, dated 1545, 245 mm in diameter, 15 mm thick and it weighs 3.97 kg, currently on display in the Corpus Christi Museum, Texas.[93] All these specimens are Portuguese, of the thick-wheeled type with close alidade.

For even better wind resistance, the dimensions were limited to no more than five or six inches in diameter, although this reduced the accuracy of the reading. To improve accuracy of measurements and to prevent faulty graduations or incorrect suspension, the instrument had two 90° dials, so that the navigator could take two sequential measurements by turning the alidade upside down. The instrument was ballasted by weighing down the lower spoke of the wheel towards the middle of the 16th century for even more wind resistance.[94]

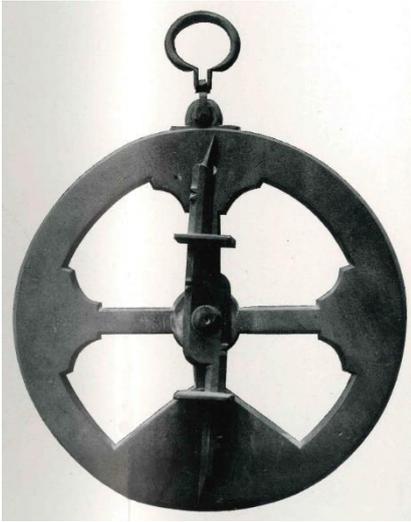

**Fig. 8.** *The nautical astrolabe of the National Museum of Palermo, lost after the Second World War (from* Coelum*).*

The nautical astrolabe was mainly used to observe the Sun and to this purpose, the holes in the pinnules were very small.[95] The user held it suspended low in front of his face inserting his thumb, or rather a string or a wire, into the upper ring.[96] Turning the movable arm caused the spot of light produced by the upper hole to fall on the lower hole. The altitude was read on the graduated scale on the edge. The instruments that also observed the stars were equipped with a further pair of holes on the alidade but larger in size, through which the star was observed keeping it carefully in the centre of the holes.[97] For the observation of stars, the astrolabe had to be held suspended high in front of the user, but for accurate measurement three operators were needed, one to hold the astrolabe suspended, one to align the alidade on the stars, and one to read the index.[98] The instrument was either made of bronze or brass.[99]

It seems that the earliest astrolabes used at sea were very inaccurate (see Mestre João's letter in note 82) but the development of the brass wheel model turned the astrolabe into a much better instrument. Even so, sailors preferred to go ashore if they had the chance, where they could achieve an accuracy of around half a degree.[100] The nautical astrolabe was used in England until 1657, in France until 1690, in Holland until 1675, and in Spain until 1702.[101]

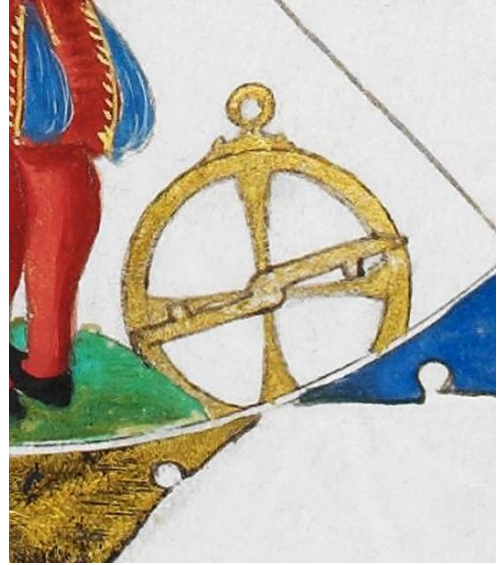

**Fig. 9.** *The astrolabe in the* Book of Hydrography *by Jean Rotz.*

## The nautical cross-staff

The cross-staff was first described in the West by the French astronomer of Jewish origin Levi Ben Gerson in 1342,[102] but the instrument was known in China as early as the 11th century and may have been introduced into Europe through Arab and Jewish circles.[103] Its invention may also have been inspired by the *kamal*, as the operating principle is similar.[104]

Gerson's instrument consisted of two pieces of wood arranged in such a way that the shorter piece could slide along the longer piece. The eye was at one end of the longer graduated piece (the staff) and the shorter piece (or crossbar) was then moved so that a star could be seen at each end of the shorter piece. The observer

---

could then determine the angular separation between the two stars by reading it on the graduations of the staff.[105]

Regiomontanus was familiar with the treatise in which Gerson described the instrument and wrote a brief description of it himself.[106] His patron Bernhard Walter used it for astronomical observations in Nuremberg from 1475 to 1488.[107] In the description by Johann Schöner, who was probably the first to call it *radius astro-nomicus*,[108] the staff was subdivided into linear, not angular units, and trigonometric calculations were therefore necessary to find the angle. But Johannes Werner, in 1514, had  introduced the division into angles,[109] without the need for calculations, probably to facilitate sailors, since the cross-staff  was the ideal instrument for measuring longitude at sea with  the method of lunar distances , which he promoted.[110]  Werner's is also the first published illustration of the cross-staff.[111]  As an astronomical instrument it was discontinued after the end of the 16th century, because Brahe considered it inaccurate.[112] In addition to the astronomical version, one for topographical measurements, the *radius geometricus*, was also developed.[113]

The first printed nautical manual, the *Regimento do estrolabio e do quadrante* (ca. 1509) only mentions the astrolabe and the quadrant but not the cross-staff, as does Duarte Pacheco Pereira in the *Esmeraldo de*

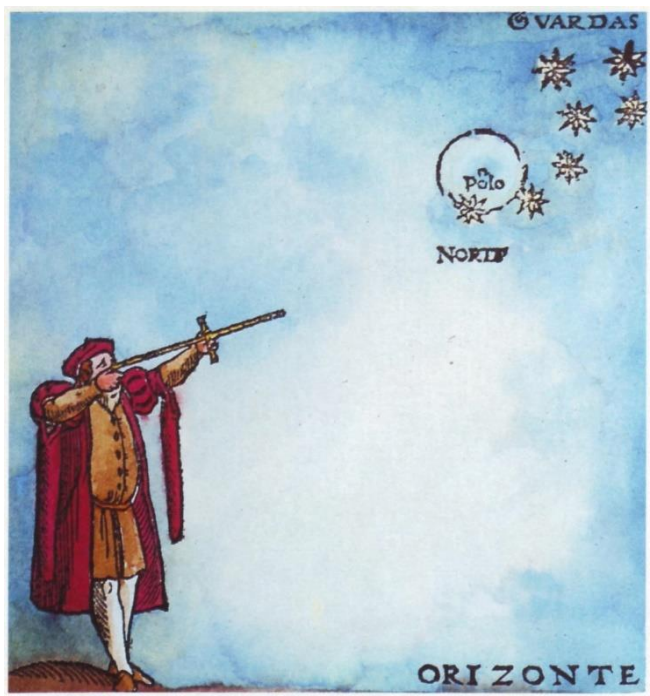

*situ orbis* (1505-1508), and Valentim Fernandez in the *Reportorio dos tempos*, 1521 and 1528 editions.[114]  The earliest references are attributed to João de Lisboa and André Pires. Both wrote their manuscripts before 1520, the former probably around 1514.[115] When taking measurements at sea the observer had to move the crossbar until the lower part touched the horizon and the upper part the star so that he could read the altitude on the graduated staff.  Although it could technically be used for the Sun, the glare would make the measurement quite difficult.  The first drawing  that shows it in use as a nautical instrument appeared in de Medina.[116] Martin Cortes, in 1551, was the first author to publish the construction details of a cross-staff.[117]

Outside the Iberian Peninsula, the nautical cross-staff spread first to France and Great Britain, between 1533 and 1540, and then around 1558 to the Netherlands, where it was greatly improved at the end of the 16th century.[118] Being 2.5 feet, it was much shorter than the astronomical cross-staff, which was 5-6 feet long, to ensure greater manoeuvrability on board ships.[119] Initially it only had one crossbar and it was not until 1581 that a version with three crossbars was developed. Of the crossbars, one usually measured angles from 10° to 30°, one measured angles from 20° to 60° and the third angles from 30° to 90°, on three different scales; the shortest one could also be used for width to measure angles from 4° to 10°.[120] The crossbars were 15, 10 and 6 inches long.[121] It was not until about 1650 that the version with four crossbars and four different scales was developed.[122] A little later a standard length of the staff of three feet with a three-quarter-inch section and three crossbars became established in Britain,[123] while the Dutch model had an average length of 800 mm and a width of 16 mm, with 40 mm wide flat crossbars.[124]

*Fig. 10. A cross-staff observation of Polaris. The circle the star describes at night around the pole is also visible (from de Medina*, Regimiento de navegación*).*

Although the use of the cross-staff was simpler than the use of the quadrant and the astrolabe, obtaining accurate results was just as complicated. Keeping a star and the horizon in sight at the same time was a difficult skill to acquire. A further complication, equally affecting the use of the quadrant and astrolabe, was the need to keep the instrument stable and vertical on the swaying deck of a ship. The parallax error, the fact that the observer cannot always keep the end of the staff in the correct position against his cheekbone, would also impact on accuracy. Moreover, the cross-staff could not be practically used for stars over 50° in altitude both because this value is close to the limit of an observer's field of view and because, at least until the advent of the three-crossbar model, the graduations on the staff became so close together that the slightest observation error would result in an error of several degrees.[125]

However, it seems that it could provide results as accurate as the quadrant and the astrolabe once the user had mastered its use: recent measurements at sea have shown an accuracy of the order of a third of a degree.[126] Whereas the astrolabe is certainly easier to use with the Sun, the cross-staff is more conveniently used with the Polaris, and in this latter respect it performs even better than the quadrant.

The oldest known nautical cross-staff is the specimen belonging to the Dutch navigators Jacob van Heemskirck and Willem Barentszoon. Abandoned in Novaya Zemlya in 1597, it was found in an expedition in 1876 and is now housed in the Rijksmuseum in Amsterdam.[127]

**Techniques**

**Polar altitude**

According to Poulle, it was Raymond de Marseille, who around 1140, first in the West, suggested that the measurement of latitude be based on the observation of the altitude of stars, whereas Masha'allah ibn Atharī seems to have been the first to suggest it at all (ca. 740-815). Moreover, Raymond de Marseille's treatise on the astrolabe, the *Liber cursuum planetarum*, seems to bear the earliest reference to the conduct of ships. He recommends double observation of a circumpolar star at the upper and lower culmination. In the course of the 12th century, other treatises specified that the maximum altitude of any star with known declination could be used, also because double observation is in fact only possible during long winter nights.[128]

In the Middle Ages, Polaris had come so close to the pole that it could be used directly for altitude measurements, i.e. its altitude above the horizon was approximately equal to the latitude of the spot from which the measurement was taken. From 1217 onwards its distance became less than 5°. As a matter of fact several sources can be found from that time which, especially with reference to the compass, mention it as a means of finding the north. Thus Alexander Neckam in his *De utensilibus* around 1190:

…and the sailors will thus know how to direct their course when the pole star is concealed through the troubled state of the atmosphere.[129]

As the French trouvère Guiot de Provins puts it in the satirical poem *Bible*, composed around 1204:

*Then the point turns direct*
*to the star with such certainty*
*that no sailor will ever doubt it,…[130]*

From Jacques de Vitry:

The iron needle, after touching the diamond, always points towards the northern star, which, as the axis of the firmament, does not move while the others decline.[131]

From the Spanish code of laws *Siete Partidas*, compiled between 1256 and 1265 under Alfonso X of Castile:

Just as mariners are guided during the night by the needle, which replaces for them the shores and pole star alike,…[132]

Brunetto Latini, *Li Livres dou Trésor*, written between 1261 and 1268:

For this reason sailors sail at the sight of those stars, which they call tramontane, and those who are in Europe and in these parts sail at that north tramontana, and the others at that of noon.[133]

*Mare Amoroso*, an anonymous Tuscan poem written around 1270-80 (vv. 293-294):

*…sì com' lo marinaro vène a porto*
*guidandosi per l'alta tramontana…*

Monte Andrea, a Guittonian poet, in a sonnet written around the same time:

*Sì come i marinar' guida la stella,*
*che per lei ciascun prende suo vïagio,…*

Marco Polo gives a report about Sumatra, "… that this island is so far to the south that the North Star is never to be seen there, neither little nor much,"[134] while "something" can be seen from Cape Comorin and even "more" from Malabar, Gujarat and Cambay. Sailing from Cape Comorin 30 miles to the north the star could be seen one cubit high on the sea horizon, two cubits from Malabar, six cubits from Gujarat.[135]
These lines from Dante:

*…si mosse voce, che l'ago alla stella*
*parer mi fece in volgermi al suo dove;…[136]*

identify the position of the star with the one indicated by the compass needle; and:

*…imagini la bocca di quel corno*

*che si comincia in punta dello stelo*
*a cui la prima rota va dintorno,...*[137]

where Ursa Minor is compared to a horn whose tip, identified with Polaris, coincides with the end of the celestial axis around which the sky turns.

Of course scholars knew that Polaris was not exactly on the pole, but described a small circle around it. This is what Pierre de Maricourt wrote as early as 1269, again in reference to the compass needle:

…one end will turn to the star which has been called the Sailor's star because it is near the pole; the truth is, how-ever, that it does not point to the star but to the pole itself.

Another necessary consequence of this is that the needle does not point to the pole star, since the meridians do not intersect in that star but in the poles of the world. In every region, the pole star is always found outside the meridian except twice in each complete revolution of the heavens.[138]

At the end of the 15th century, a set of rules, the so-called *Regimento do Norte*, became available to mariners for consultation. It supplied corrections to be applied to the altitude of Polaris in the various positions to determine latitude. They were probably circulating in hand-written form as early as 1480-81 as part of a complete guide to nautical rules,[139] which was probably printed around 1495[140] and whose first known edition probably dates from 1509.[141] The guide was known as the aforementioned *Regimento do estrolabio e do quadrante*,[142] and it includes a translation of Sacrobosco's *Tractatus de Sphaera* intended to give pilots the basics of astronomy; two versions of the regiment for the determination of latitude from the Sun's meridian altitude and declination; a table with solar declination; the regiment for the determination of latitude from the altitude of the pole star; a wheel with the altitudes of the pole star from Lisbon.[143] It is laid out in very simple form, with many examples, to be understandable to sailors. Around the same time the *Regimento do Norte* was available in the *Reportório dos Tempos* published by Pablo Harus in Zaragoza in 1492 and later translated into Portuguese and published by Valentim Fernandes in 1518.[144]

The *Regimento do Norte* was first drawn up in the form of a wheel with the altitudes of the North Star from the port of departure, e.g. Lisbon, according to eight different positions of the Guards, Kochab and Pherkad, stars that sailors knew well because they were used to measure the night hours by means of an instrument called nocturnal, or even by sight. Thus a different wheel was needed for each place the pilot wanted to visit. The next step was to publish correction values valid for any location. For example, when the Guards were exactly at the top one had to add 3°, when they were exactly at the bottom one had to remove them.

Originally, the pilot was taught to use his quadrant as a means of measuring his linear distance south or north of his port of departure, usually Lisbon. He was taught to observe the altitude of the North Star at his port of departure when the Guards were in a given position and to mark the altitude on the quadrant scale, as indicated by the plumb line. Later, during the voyage, he would observe the North Star when the Guards were in the same position and mark where on the scale the plumb line cut it. In the mid-fifteenth century, on the route from the African coast to the Azores or Madeira and vice versa, it became common practice to mark the dial with the altitude of the North Star (with the Guards in a given position) at various points along the coast or on the islands, and pilots would then sail along the coast or in the open sea to the designated

---

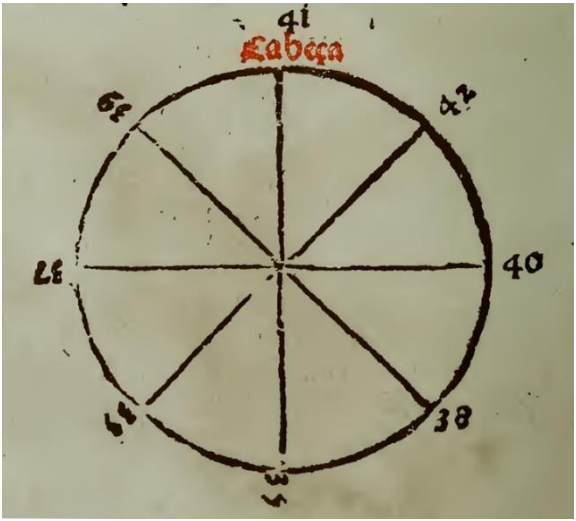

**Fig. 11.** *The Regimento do norte to find the altitude of the pole according to the position of the Guards and of Polaris (from* Regimento do estrolabio e do quadrante*).*

altitude and then set the bow straight for east or west to reach their landing place (navigation in latitude).[145] Only later, as the pilots' observational and mathematical skills improved, did they learn how to directly convert the altitude measurement into observed latitude by means of the North Star rule.

For polar altitude measurements the quadrant is more convenient than the astrolabe, because the use of the quarter circle instead of the full circle doubles the accuracy of the measurement. If one is using an astrolabe instead, it is difficult to sight the celestial body through the alidade while keeping the instrument suspended at the same time, unless one suspends it from the mast or resorts to an assistant, and even the procedure is complicated.

Although, as mentioned above, any star could be used, at that time sailors tended to use the Polaris for a number of reasons. Firstly because, unlike the other stars, it could be observed at any time by means of the *Regimento do Norte*; secondly, its position was the easiest to spot; thirdly, it was much more difficult to capture the exact moment of th e meridian transit of the other stars; finally, even later with an instrument such as the cross-staff, which requires the simultaneous visibility of both the horizon and the star, it was not certain that a star of sufficient brightness known to the sailor would pass in the meridian in that interval of about half an hour called nautical twilight, whereas the Polaris was always there.[146]

**Altitude of the Sun**

At the end of the fifteenth century, continuous and effective observation of the Polaris proved difficult, already at about 8-9° N latitude, owing to atmospheric extinction and the heaving of the ship, except in its eastern elongations: in its lower transit it would be completely invisible, while in its upper and western elongations the Guards would be invisible. Therefore, in the following years, the Portuguese trained themselves to the method of measuring latitude based on the meridian altitudes of the Sun.

According to Poulle, the first tables of declination of the Sun in the West also appeared around 1140 in the astrolabe treatise *Liber cursuum planetarum* by Raymond de Marseille, and they were taken from the *Toledan Tables*, compiled around 1080.[147]

The method was developed to the benefit of mariners by the Jewish astronomer Abraham Zacuto between 1473 and 1478, published in his *Almanach perpetuum*,[148] and simplified and verified by two other Jewish astronomers in the service of Portugal, José Vizinho (who translated the *Almanach* into Latin),[149] and Mestre Rodrigo who, around 1485, travelled along the coast of Guinea to test its validity and establish the latitudes of the various ports and key docking points.[150] This method is of course not as easily applied as the other. Tables would show the daily declination of the Sun, i.e. its distance from the celestial equator. The sailor would have to measure the altitude of the Sun at noon, when it appeared highest. To be sure not to miss the moment of culmination, the measurement had to start some time before the culmination and continue for a while afterwards. The best device for doing this was the astrolabe, because with the quadrant you would risk losing your eye. A set of three simple algebraic operations, convenient in the various cases where the Sun was north or south of the equator and north or south of the zenith, made it possible to find latitude. In time, the scale of heights on the instrument was replaced by the scale of zenith distances, thus simplifying the algebraic operation.

A *Regimento do sul*, based on the Southern Cross, was also submitted by João de Lisboa in 1514 in his *Tratado de agulha de marear*[151] but, as the constellation was at a considerable distance from the celestial south pole, it was very approximate and its use was rather limited.[152] The rule used the stars of the Southern Cross itself as "guards" or pointers. Thus, for example, when the head and foot of the Cross were in line with a plumb line held in front of the navigator, the foot was exactly 30° above the celestial south pole.[153]

**The accuracy of early latitude measurements**

Unfortunately, not much quantitative data on the first latitude measurements are left. We do not even have the raw measurement of those of Gomes and Azambuja. As for Diaz's, his measurement of the latitude of the Cape of Good Hope, 45° S,[154] was wrong by almost 11°.

Evidence from Colombo's notes proves that he had both astrolabes and quadrants on board during his first voyage (1492-93),[155] but only used the latter, obtaining contradictory and apparently absurd values[156] that were to puzzle many historians.[157] Colombo also allegedly knew how to derive latitude from the meridian altitude of the Sun: he claimed to have done so on the voyages from Lisbon to the Guinea coast when he was in Portugal,[158] around 1480, but there is no reference in his writings to measurements made by this method during the four discovery voyages.

According to the only historian[159] who provides a quantitative indication of the measurement Da Gama took with the large wooden astrolabe at St. Helena Bay, he appears to have thought he was at most 30 leagues from the Cape, which means, taking the value adopted by the Portuguese at that time for the degree of latitude, 17.5 leagues,[160] 1°43' difference, whereas the real value is 1°36'. This seems a reliable finding, but *at most 30 leagues* does not mean *at 30 leagues*, so the error may have been greater, and that is assuming Da Gama knew what the exact latitude of the Cape was!

In his third voyage,[161] in July 1498, with the Polaris being only between 7° and 13° high, and therefore difficult to observe, the six measurements made by Colombo were affected by an average error of 2°16'.[162]

Mestre João made an error of two thirds of a degree in measuring the latitude of the bay of Coroa Vermelha, where Cabral's fleet had landed in Brazil.[163]

| Location | Albo | Pigafetta | Correct |
|---|---|---|---|
| Tenerife | | 28°00′ N | 28°19′ N |
| Cabo Frio (Brazil) | 23° S | | 22°53′ S |
| Cabo Santa Maria (Uruguay) | 35° S | | 34°40′ S |
| Mouth Rio Santa Cruz (Argentina) | | 50°40′ S | 50°07′ S |
| Cabo S. Antonio (Argentina) | 36° S | | 36°40′ S |
| San Julian (Argentina) | 49°40′ S | | 49°18′ S |
| CaboVirgenes (Argentina) | 52° S | 52° S | 52°23′ S |
| Isola Limasawa (Philippines) | 9°40′ N | 9°40′ N | 9°56′ N |
| PanglaoIsland (Philippines) | 9°20′ N | | 9°35′ N |
| Cebu (Philippines) | 10°20′ N | | 10°18′ N |
| Bohol Island (Philippines) | 9°30′ N | | 9°50′ N |
| MouthQuipitRiver (Mindanao) | | 8° N | 8°04′ N |
| Cagayan Sulu Island (Philippines) | | 7°30′ N | 7°01′ N |
| Puerto Princesa (Palawan) | | 9°20′ N | 9°44′ N |
| Bandar Seri Begawan | 5°25′ N | 5°15′ N | 4°56′ N |
| Jolo Island (Philippines) | 6° N | | 5°58′ N |
| BasilanIsland (Philippines) | 6°50′ N | | 6°34′ N |
| SaranganiIsland (Moluccas) | 4°40′ N | | 5°27′ N |
| SangiheIsland (north-east of Celebes) | 3°40′ N | 3°30′ N | 3°35′ N |
| SiauIsland (Moluccas) | 3° N | | 2°42′ N |
| Tidore Island (Moluccas) | | 0°27′ N | 0°40′ N |
| Ternate Island (Moluccas) | 1° N | 0°40′ N | 0°48′ N |
| BacanIsland (Moluccas) | | 1° S | 0°35′ S |
| BuruIsland (Moluccas) | 3°30′ S | 3°30′ S | 3°24′ S |
| AlorIsland (SundaIslands) | 8°20′ S | 8°30′ S | 8°15′ S |
| Cape of Good Hope | | 34°30′ S | 34°24′ S |
| Mouth Casamance River (Senegal) | 12°03′ N | | 12°33′ N |

**Tab. 1.** *Some latitudes from the Reports of Albo and Pigafetta compared to modern values.*

Columbus, in the only latitude observation recorded on his fourth voyage from the Bay of St. Gloria (now St. Anne) in Jamaica, committed an error of less than half a degree, the result though of the average of several measurements made over an entire year, between 1503 and 1504, from the deck of a stranded ship.[164]

The first fairly accurate and repeated assessments of latitudes at sea were made during Magellan's voyage. The Portuguese navigator took 21 quadrants and seven astrolabes on board his ships, which showed his great interest in the problems of celestial navigation.[165] In the two most detailed existing accounts of the enterprise coming from Pigafetta and pilot Francisco Albo,[166] several latitudes of the landings are reported. They are almost always correct to within half a degree and are the result of recording, whenever possible, the meridian altitudes of the Sun or, more rarely of the Polaris, given that navigation almost always took place below 9° N. That is a reasonably successful achievement, if we compare some of these data randomly (tab. 1) to some latitude measurements reported in Regiomontanus' *Calendarium* (tab. 2).[167] The average error of the latter is 55'. If we confine ourselves to the data of the cities where Regiomontanus worked (Nuremberg, Buda, Rome and Vienna) and we assume that he measured them directly, the average error is 19'. Pigafetta's and Albo 's errors are 16' and 18' respectively, i.e. in the same period the best pilots seemed to be as competent as the best astronomers, and even better.

---

[164] *Libro de las profecias*, in *Nuova Raccolta Colombiana* (Rome: Istituto Poligrafico e Zecca dello Stato, 1993), 3(1) p. 195.

[165] F.H.H. Guillemard, *The life of Ferdinand Magellan* (London: Philip & Son, 1890), pp. 334-335.

[166] Pigafetta, *op. cit.* (note 1), pp. 51-112; *Diario ó derrotero del viage de Magallanes desde el cabo de San Agustín en el Brasil, hasta el regreso á España de la nao Victoria, escrito por Francisco Albo*, in Martin Fernandez de Navarrete, *Coleccion de los viages y descubrimientos que hicieron por mar los españoles desde fines del siglo XV* (Madrid: Imprenta Nacional, 1837), 4, pp. 209-247.

| City | Regiomontanus | Correct |
|---|---|---|
| Lisbon | 41° | 38°43' |
| Paris | 48° | 48°52' |
| Toledo | 41° | 39°52' |
| Toulouse | 43° | 43°36' |
| Salzburg | 48° | 47°48' |
| Krakow | 51° | 50°03' |
| Nuremberg | 49° | 49°27' |
| Rome | 42° | 41°54' |
| Buda | 47° | 47°30' |
| Magdeburg | 54° | 52°08' |
| Marseille | 43° | 43°18' |
| Vienna | 48° | 48°13' |
| Regensburg | 49° | 49°01' |
| Prague | 50° | 50°05' |
| Ulm | 47° | 48°24' |
| Milan | 44° | 45°28' |
| Leipzig | 51° | 51°19' |
| Mainz | 50° | 50°00' |
| Augsburg | 46° | 48°22' |
| Venice | 45° | 45°26' |
| Naples | 43° | 40°50' |
| Genoa | 43° | 44°24' |
| Ancona | 44° | 43°37' |
| Lübeck | 56° | 53°52' |
| Florence | 43° | 43°44' |

**Tab. 2.** *The latitudes of some cities from the* Calendarium *of Regiomontanus compared to modern values.*

In conclusion, before even the nautical cross-staff came into use, the first phase of celestial navigation aiming at measuring latitude at sea with sufficient precision, came to an end 500 years ago, having begun several centuries earlier in various parts of the world. As we have seen, the idea had built up in ancient times, when there were instruments suitable for the purpose, originally designed for astronomical observations, but practical implementation took a long time and only came into being at the beginning of the modern age, when instruments and rules were simplified to the point of being accessible to sailors.

---

[167] Johannes Regiomontanus, *Calendarium* (Venice: Ratdolt, 1483), Tabula regionum.